\documentclass[fp,twocolumn]{jpsj3}
\usepackage{txfonts}
\usepackage[dvipdfmx]{graphicx}%
\usepackage{amsmath}%
\usepackage{amsfonts}%
\usepackage{amssymb}
\usepackage{bm}
\usepackage{color}
\usepackage{ulem}

\def\e{{\epsilon}}
\def\k{{ {\bm k} }}

\def\q{{ {\bm q} }}
\def\Q{{ {\bm Q} }}
\def\w{{\omega}}
\def\a{{\alpha}}

\title{Non-Fermi-Liquid Transport Phenomena in Infinite-Layer Nickelates}

\author{Shinichi Hiragami and Seiichiro Onari\thanks{onari@s.phys.nagoya-u.ac.jp}}
\inst{ Department of Physics, Nagoya University, Furo-cho, Nagoya 464-8602, Japan.} 

\abst{Recently discovered superconducting infinite-layer nickelates
   $R$NiO$_2$ ($R$=Nd, La, Pr) attract increasing attention due to their
   similarities to cuprates. Both $R$NiO$_2$ and YBCO cuprates exhibit
   the non-Fermi-liquid transport behavior, characterized by resistivity
   proportional to temperature near the quantum critical point of the
   charge or spin density wave. In this study, we analyze the
   resistivity of infinite-layer nickelate Nd$_{0.85}$Sr$_{0.15}$NiO$_2$
   based on a three-dimensional tight-binding model within the framework of the quasi-particle picture by applying linear response theory. We take account of the self-energy by the fluctuation-exchange approximation for the Ni orbital and the T-matrix approximation for an impurity effect on the Nd orbitals. We find that (i) the $T$-linear resistivity at low temperatures is derived from the spin fluctuations, and (ii) a negative and $T$-linear Seebeck coefficient is obtained. 
   Therefore, NdNiO$_2$ behaves as a quasi-two-dimensional electron system, similar to CeCoIn$_5$.
   }

\begin{document}
\maketitle

\section{Introduction}
The superconducting nickelates have recently stimulated much attention
 due to their similarities in electronic structure and phase diagrams to those of cuprates.
In particular, the infinite-layer nickelates $R$NiO$_2$ ($R$=Nd, La, Pr)
 with the superconducting transition temperature $T_c \gtrsim 10$K
 \cite{Ni-super1,Ni-super2,Ni-super3,Ni-super4} possess a Ni-$3d^9$ configuration, analogous to the Cu-$3d^9$ configuration in cuprates.
On the other hand, high $T_c\sim 80$K has been observed in the
 double-layer nickelate La$_3$Ni$_2$O$_7$ under pressure\cite{2Ni-super}.
 The double-layer nickelate possesses a Ni-$3d^{7.5}$ configuration.
 In these nickelates, the superconducting state emerges near the quantum
 critical point (QCP) of charge density wave (CDW) state, which is also similar
 to cuprates.
However, a significant difference between infinite-layer nickelates and cuprates
   is the dimensionality of the Fermi surface (FS).
The FS in $R$NiO$_2$ exhibits three-dimensional (3D) characteristics, in contrast to the two-dimensional (2D) FS in cuprates.

Recently, resonant inelastic X-ray scattering (RIXS) measurements have
detected a CDW in $R_{1-x}$Sr$_x$NiO$_2$
\cite{Ni-CDW1,Ni-CDW2,Ni-CDW3,Ni-CDW4}.
The 3D structure of the CDW exhibits a period 3 along the
$x$ direction and a period 3 to 5 along the $z$ direction, which has been explained by our previous
theoretical work based on the paramagnon-interference mechanism
\cite{Ni-Onari}. The QCP of the CDW appears at $x\sim 0.15$ \cite{Ni-CDW1}.
In contrast, very recently, the absence of period 3 along the
$x$ direction has been reported \cite{no-period3}. Thus, the 3D structure of the CDW has yet to
be solved.
On the other hand, the spin fluctuations are also developed in $R_{1-x}$Sr$_x$NiO$_2$, as indicated by a moderate increase in $1/T_1T$ for $T\lesssim 100$K \cite{Ni-NMR}.
 
In proximity to the QCP of the spin density wave (SDW) or CDW, non-Fermi-liquid transport
 phenomena have been observed in both nickelates and cuprates
 \cite{Ni-Non-Fermi1,Ni-Non-Fermi2,Ni-Non-Fermi3,Non-Fermi-Cu,Cu-trans1,Cu-trans2,Non-Fermi}.
 In $R_{1-x}$Sr$_x$NiO$_2$ near the QCP of the CDW, the non-Fermi-liquid transport
 phenomena such as a resistivity $\rho\propto T$ and
 a $T$-dependent Hall
 coefficient have been observed \cite{Ni-Non-Fermi1,Ni-Non-Fermi2,Ni-Non-Fermi3}. 
 In triple-layer and quintuple-layer nickelates, a negative and 
 $T$-linear Seebeck coefficient has been observed \cite{Ni-S}.
 The origin of the non-Fermi-liquid transport phenomena in nickelates is
 an unsolved problem.
 
 Non-Fermi-liquid behaviors of cuprates have been explained by the spin
 fluctuation theories, such as the self-consistent renormalization (SCR) theory \cite{moriya-takahashi,ueda-moriya,moriya-ueda} and
 the fluctuation-exchange (FLEX) theory \cite{bickers1,bickers2,monthoux}.
 The spin-fluctuation theories\cite{moriya-ueda,stojkovic}
derive the relations $\rho\propto T$ in 2D systems and $\rho\propto T^{1.5}$ in
 3D systems near the QCP of the SDW. In fact, $\rho\propto T$ has been
 observed in cuprates \cite{Non-Fermi-Cu,Cu-trans1,Cu-trans2}, which are 2D systems.
  In addition, Kontani \cite{Non-Fermi} has developed
 the spin fluctuation theory considering the current vertex corrections,
 which play an important role for the Hall coefficient and the Nernst
 coefficient in cuprates.
 We note that the relation $\rho\propto T$ observed in CeCoIn$_5$
 \cite{Ce-trans} has
 also been explained by the spin-fluctuation theory \cite{Onari-Ce}.
Despite having the 3D FS and 3D bandstructure, CeCoIn$_5$ exhibits 2D transport phenomena. This means that the electronic state of CeCoIn$_5$ is quasi-2D, which defines the usage of the term “quasi-2D” in this paper.

In this paper, we study the origin of non-Fermi-liquid $T$-linear resistivity in
infinite-layer nickelate Nd$_{0.85}$Sr$_{0.15}$NiO$_2$ based on a 3D
tight-binding model within the framework of the quasi-particle picture. We employ the FLEX approximation
for the
Ni $d_{x^2-y^2}$ orbital and the T-matrix approximation for an impurity
effect on the Nd $d_{z^2}$
and $d_{xy}$ orbitals. The CDW fluctuations are ignored for simplicity
in the present study.
We find that the 3D spin fluctuations induce $T$-linear resistivity at low temperatures.
Thus, this system is quasi-2D.
We note that the obtained slope of
the resistivity with respect to $T$ is slightly smaller than
experiments. This slope is expected to increase and approach the experimental value when the CDW fluctuations are taken into account.
We also find that the obtained Seebeck coefficient is negative and 
$T$-linear, which is consistent with experimental results in
triple-layer and quintuple-layer nickelates \cite{Ni-S}.

\section{Model and Hamiltonian}
Since the non-Fermi-liquid $T$-linear resistivity has been observed
around the hole doping $x=0.15$ in Nd$_{1-x}$Sr$_x$NiO$_2$,
we analyze the following 3D
three-orbital Hubbard model for $x=0.15$, where $d_{x^2-y^2}$ orbital of Ni,
$d_{z^2}$ and $d_{xy}$ orbitals of Nd are taken into account:
\begin{equation}
H=H^0+H^U, 
\end{equation}
where $H^0$ is the tight-binding model employed on
Ref. \cite{Ni-Onari}. We introduce the energy shift $-0.25$eV of Ni $d_{x^2-y^2}$ orbital
in order to reproduce
the bandstructure with small Fermi pocket of Nd orbitals obtained by the
density functional theory (DFT)+DMFT calculation.
\cite{Ni-first-principle6}.
Details of the present model are explained in the Appendix.
Figures \ref{fig1}(a) and (b) show
band dispersion and 3D FSs in this model. 
Orbitals $1$, $2$, and $3$ denote Ni $d_{x^2-y^2}$ orbital,
Nd $d_{z^2}$ orbital, and Nd $d_{xy}$ orbital, respectively.
We stress that an analysis using the 3D model is essential for this
system, since the FSs have a 3D structure.
$H_U$ is the Coulomb interaction, where the Coulomb
interaction of only
orbital 1 is taken into account, since the results, including the Coulomb
interactions of the three orbitals, are almost the same.

\begin{figure}[!htb]
\includegraphics[width=.8\linewidth]{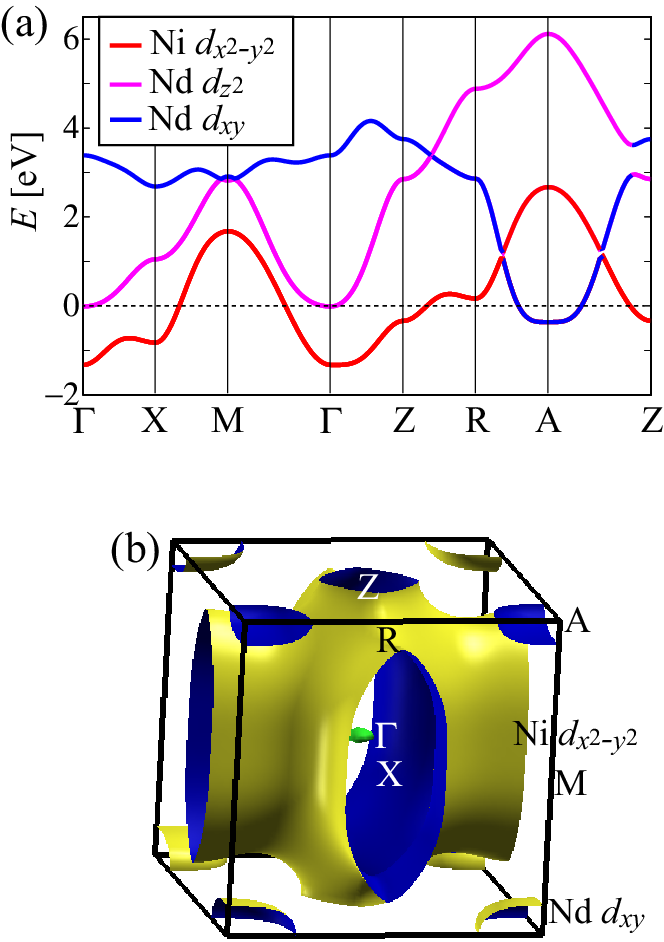}
\caption{
 (a) Bandstructure of the present Nd$_{0.85}$Sr$_{0.15}$NiO$_2$ model. Red, purple, and blue
 lines represent Ni $d_{x^2-y^2}$, Nd $d_{z^2}$, and Nd $d_{xy}$
 orbitals, respectively.
  (b) 3D FSs in the present model.}
\label{fig1}
\end{figure}

 \section{Formulation}
First, we analyze the self-energy of orbital $1$ in the FLEX approximation.
The self-energy in the FLEX approximation $\hat{\Sigma}^{\rm FLEX}$ with
only ${\Sigma}^{\rm{FLEX}}_{1,1}$ component, the effective interaction
for the self-energy in the FLEX approximation $V^{\rm{FLEX}}$, the spin
(charge) susceptibility $\chi^{s(c)}$, the irreducible susceptibility
$\chi^0$, and Green's function $G_{1,1}$ of orbital $1$ are given as
\begin{eqnarray}
{\Sigma}^{\rm{FLEX}}_{1,1}(k)&=&\frac{T}{N}\sum_{q}V^{\rm FLEX}(q)G_{1,1}(k-q) \label{Eq1}\\
  V^{\rm{FLEX}}(q)&=&\frac{3}{2}U_1^2\chi^s(q)+\frac{1}{2}U_1^2\chi^c(q)-U_1^2\chi^0(q) \\
{\chi}^{s(c)}(q)&=&\frac{{\chi}^0(q)}{1-(+)U_1{\chi}^0(q)} \\
\chi^0(q)&=&-\frac{T}{N}\sum_kG_{1,1}(k+q)G_{1,1}(k) \\
G_{1,1}(k)&=&\left[\frac{1}{(i\e_n+\mu)\hat{1}-\hat{h}^0(\k)-\hat{\Sigma}^{\rm{FLEX}}(k)}\right]_{1,1}, \label{Eq5}
\end{eqnarray}
where $k=[\k,\e_n=(2n+1)\pi T]$, $q=(\q,\w_m=2m\pi T)$. In the present
3D calculation, we employ
$N=64\times64\times64$ $\k(\q)$ meshes and 512 Matsubara frequencies. ${\hat{h}}^0(\k)$ is the matrix expression of $H^0$, 
and $\mu$ is the chemical potential.
We solve Eqs. (\ref{Eq1})-(\ref{Eq5}) selfconsistently.

Next, we consider the nonmagnetic impurity effect induced by the
substitution of Nd and Sr atoms. The impurity effect is taken into
account by the T-matrix approximation in the orbital representation;
\begin{eqnarray}
 \hat{T}(i\e_n)&=&\left[\hat{1}-\hat{I}\hat{G}^{\text{loc}}(i\e_n)\right]^{-1}\hat{I}, \\
\hat{G}^{\rm{loc}}(i\e_n)&=&\frac{1}{N}\sum_{\bm{k}}\hat{G}_{\bm{k}}(i\e_n),\\
\hat{I}&=&\begin{pmatrix}
0&0&0\\
0&I_{\text{Nd}}&0\\
0&0&I_{\text{Nd}}
  \end{pmatrix},\\
\hat{\Sigma}^{\text{imp}}&=&n_{\text{imp}}\hat{T}(i\e_n),
\end{eqnarray}
where $\hat{G}^{\rm loc}$ is the local Green's function, $\hat{I}$ is
the impurity potential, $\hat{\Sigma}^{\rm imp}$ is the self-energy
induced by the impurity effect on Nd atom, and $n_{\rm imp}$ is the
density of impurity. We fix $n_{\rm imp}=0.15$ due to the hole doping
$x=0.15$ in the present study.
The total self-energy $\hat{\Sigma}_{\rm tot}=\hat{\Sigma}^{\rm
FLEX}+\hat{\Sigma}^{\rm imp}$ is given as
\begin{equation}
 \hat{\Sigma}_{\rm tot}(\k,i\e_n)=\begin{pmatrix}
\Sigma^{\rm FLEX}_{1,1}(\k,i\e_n)&0&0\\
0&\Sigma^{\rm imp}_{2,2}(i\e_n)&\Sigma^{\rm imp}_{2,3}(i\e_n)\\
0&\Sigma^{\rm imp}_{3,2}(i\e_n)&\Sigma^{\rm imp}_{3,3}(i\e_n)
  \end{pmatrix}.
\end{equation}
To obtain the real frequency dependence, we use the analytic
continuation by the Pade approximation,
 \begin{equation}
  \hat{\Sigma}_{\rm
tot}(\k,i\e_n)\rightarrow  \hat{\Sigma}_{\rm tot}^R(\k,\omega).
 \end{equation}

 Transport phenomena are analyzed by using the linear response
 theory.
 We calculate the resistivity $\rho$ and the Seebeck coefficient $S$ in
 the band representation.
 The conductivity $\sigma_{xx}$ is given as,
 \begin{equation}
\sigma_{xx} =  e^2 \sum_{b\bm{k}}\int\frac{d\omega}{\pi}\ 
  \left(-\frac{\partial f(\omega)}{\partial \omega}\right)
  |G_b^{R}(\bm{k},\omega)|^2(v_{\bm{k},x}^b)^2,
 \end{equation}
 where $f(\omega)=[\exp(\omega/T)+1]^{-1}$, and the Green's function in the band representation $G_b^{R}$
 $(b=1,2,3)$ is
 \begin{equation}
  {G}^R_b(\bm{k},\omega)=\frac{1}{(\omega+\mu)-{h}_b^0(\bm{k})-{\Sigma}^b_{\rm{tot}}(\bm{k},\omega)}.
\end{equation}
Here, ${h}_b^0$ and ${\Sigma}^b_{\rm{tot}}$ are the band
representation of $\hat{h}^0$ and $\hat{\Sigma}_{\rm{tot}}^R$,
respectively. $-e$ $(e>0)$ is the electron charge. The velocity is given as $v^b_{\k,x}=\frac{\partial
h^0_b(\k)}{\partial k_x}$, and the resistivity along $x$-axis is given as,
\begin{equation}
 \rho=\frac{1}{\sigma_{xx}}.
\end{equation}
The Seebeck coefficient $S$ is given by
\begin{eqnarray}
 S&=&-e\frac{L_{xx}}{T\sigma_{xx}},\\
L_{xx} &=& \sum_{b\bm{k}}\int\frac{d\omega}{\pi}\ 
  \left(-\frac{\partial f(\omega)}{\partial \omega}\right)\omega\ 
|G^{R}_b(\bm{k},\omega)|^2(v_{\bm{k},x}^b)^2.
  \end{eqnarray}
In the present study, we set the Coulomb interaction
$U_1=3$eV for orbital 1, which is consistent with the value $U_1=3$-$4$eV obtained by the
first-principles calculations
\cite{Ni-Sakakibara,Ni-first-principle2}.

In the present study, we employ the FLEX approximation
within the framework of the quasi-particle picture, where the effect of the spin
fluctuations is adequately incorporated. In contrast, the contribution of CDW
fluctuations tends to be underestimated. We note that the vertex
corrections, which are the higher-order many-body effects, significantly affect the Hall coefficient and the Nernst
coefficient. In contrast, their impact on the resistivity and the Seebeck
coefficient is minor in
cuprates and CeCoIn$_5$ \cite{Non-Fermi,Onari-Ce}. Therefore, for the sake of simplicity, we neglect vertex corrections in our analysis of the resistivity and Seebeck coefficient.

\section{Results and Discussion}
First, we present the spin susceptibility for orbital 1
${\chi}^{s}(q)$ in the FLEX approximation.
The spin Stoner factor $\a_{s}$ is defined as $\a_{s}=U_1
{\rm max}_{\q}\chi^0(\q,0)$. 
Thus, ${\chi}^{s}(q)\propto(1-\a_{s})^{-1}$, and $\a_{s}=1$ corresponds
to spin-ordered state.

\begin{figure}[!htb]
\includegraphics[width=.99\linewidth]{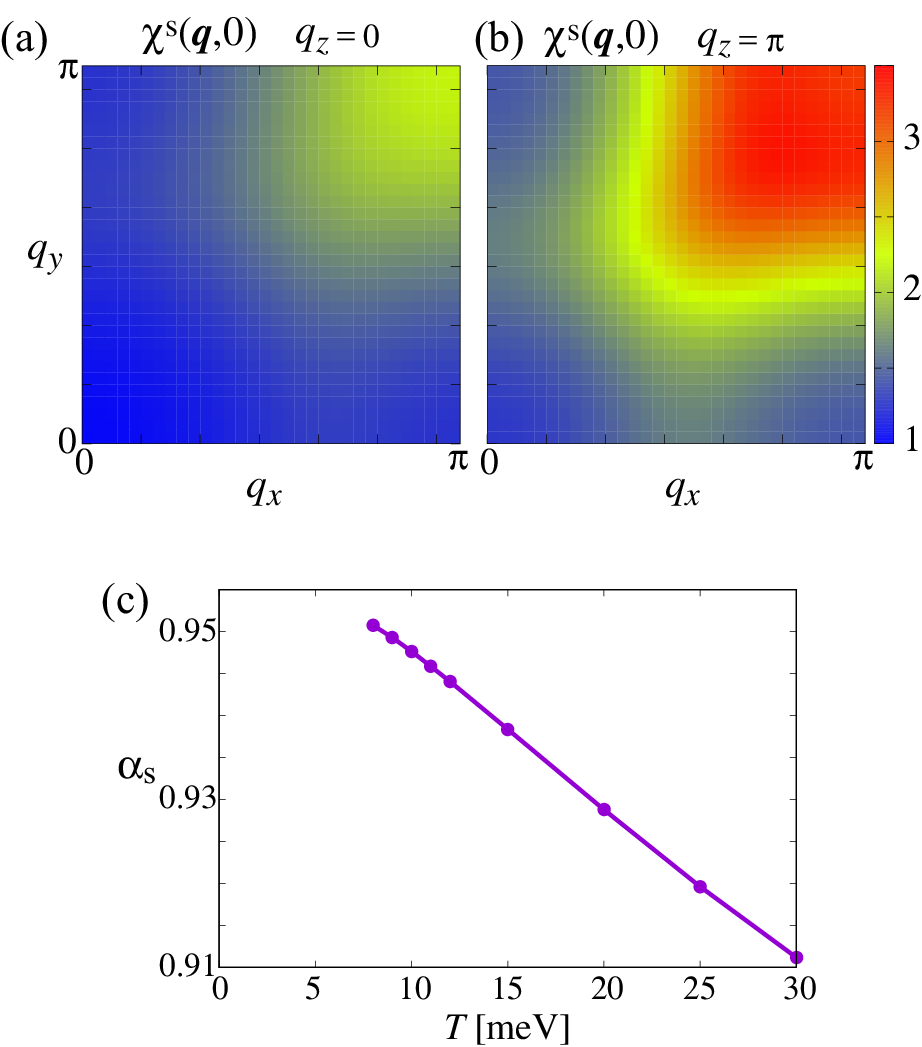}
 \caption{
(a) $\q$ dependences of 
 $\chi^{s}(\q,0)$ given by the FLEX approximation on $q_z=0$ plane, and
 (b) that on $q_z=\pi$ plane.
 (c) $T$ dependence of $\alpha_s$.
}
\label{fig2}
\end{figure}

Figures \ref{fig2}(a) and (b) show the obtained spin susceptibilities 
for orbital 1 at $T=10$meV on $q_z=0$ and $q_z=\pi$ planes, respectively.
$\chi^{s}(\q,0)$ has a broad peak around $\Q_s=(\pi,\pi,\pi)$.
These results indicate that $\chi^{s}(\q,0)$ is three-dimensional.
As shown in Fig. \ref{fig2}(c), $\alpha_s\sim 0.95$ is obtained at
$T=10$meV, which corresponds to the moderate spin fluctuations.
These moderate spin fluctuations are consistent with the experiment, where 
$1/T_1T$ moderately increases for $T\lesssim 100$K \cite{Ni-NMR}.

\begin{figure}[!htb]
\includegraphics[width=.99\linewidth]{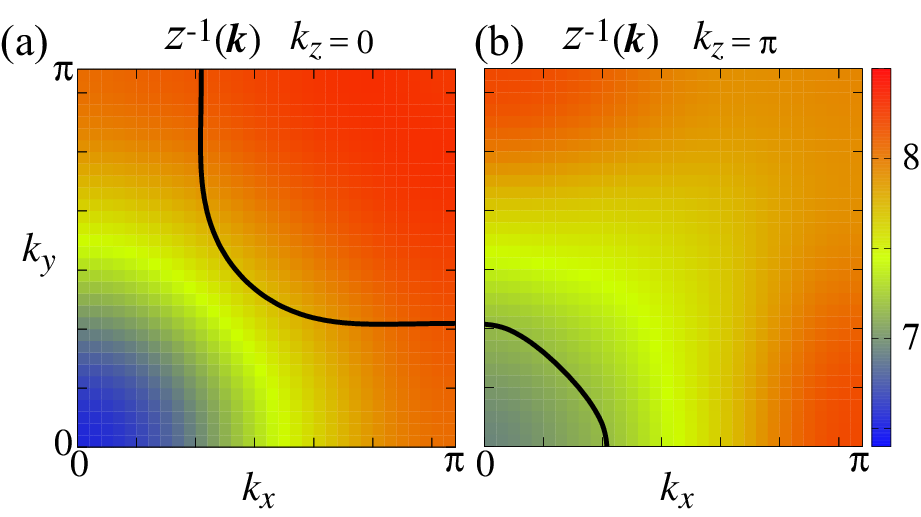}
 \caption{
(a) Obtained $\k$ dependence of mass enhancement factor $z^{-1}(\k)$ on $k_z=0$ plane, and (b)
 that on $k_z=\pi$ plane.
 Black line represents the FS.}
\label{fig3}
\end{figure}

Next, we calculate $\k$ dependence of the mass enhancement factor of orbital 1
$z^{-1}(\k)$, which is given as
\begin{eqnarray}
z^{-1}(\k)=\left[1-\frac{\partial\Sigma^{\rm FLEX}_{1,1}(\k,\omega)}{\partial\omega}\right].
\end{eqnarray}
Figures \ref{fig3}(a) and (b) show obtained $z^{-1}(\k)$ on $k_z=0$ and
$k_z=\pi$ planes, respectively. The value of $z^{-1}(\k)$ exhibits a $k_z$
dependence, reflecting the three-dimensionality of FS for orbital 1. In
addition, the
value of $z^{-1}(\k)$ on FS in
$k_z=0$ plane is
larger than that in $k_z=\pi$ plane.

Hereafter, we analyze the transport phenomena by taking account of the self-energies of both
the FLEX approximation for orbital 1 and the impurity effect for
orbitals 2 and 3.

\begin{figure}[!htb]
\includegraphics[width=.99\linewidth]{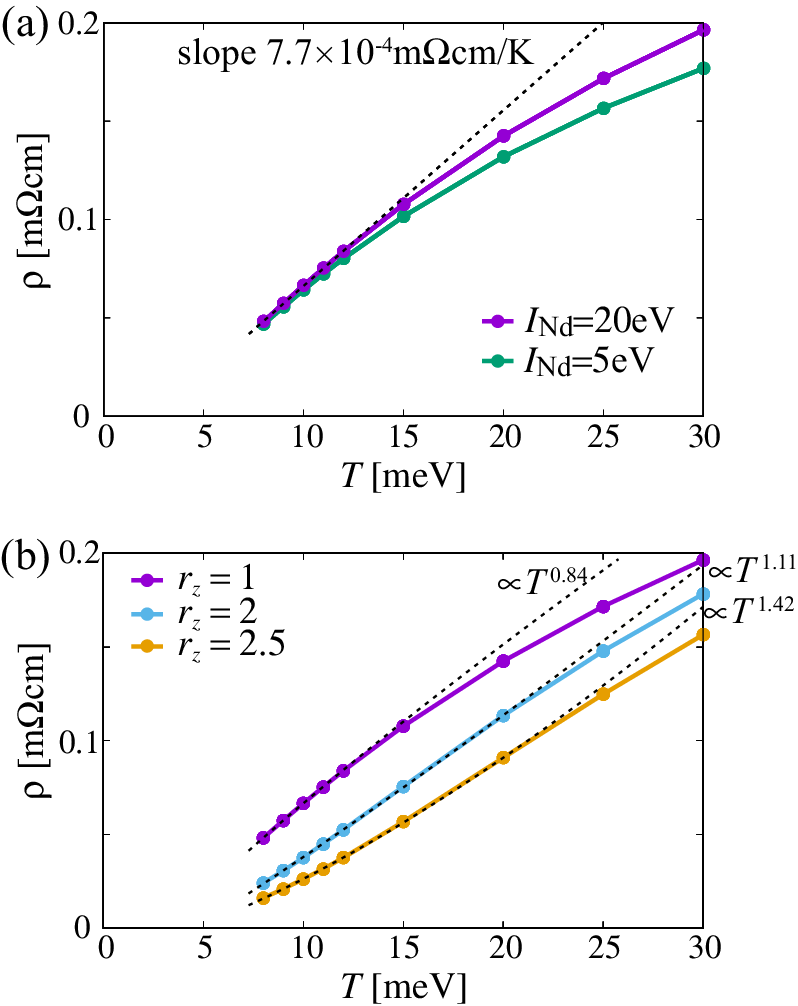}
\caption{
 (a) $T$ dependence of $\rho$ for $I_{\rm Nd}=5,20$eV.
 (b) $T$ dependence of $\rho$ for $I_{\rm Nd}=20$eV, with $r_z=1,2,2.5$.
Here, $r_z$ represents the magnification factor for the inter-layer hopping integrals.
$r_z=1$ corresponds to the original 3D model. 
}
\label{fig4}
\end{figure}

Figure \ref{fig4}(a) shows obtained $T$ dependence of resistivity $\rho$
for $I_{\rm Nd}=5,20$eV, where we use the lattice constant along $z$-axis $c=3.267$\AA. At low temperatures, $\rho$ shows $T$-linear
behavior, which is consistent with the Non-Fermi-liquid behavior
observed in experiments \cite{Ni-Non-Fermi1,Ni-Non-Fermi2,Ni-Non-Fermi3}.
The obtained slope of
the resistivity with respect to $T$, $7.7\times 10^{-4}$m$\Omega$cm/K for
$I_{\rm Nd}=20$eV, is
close to the experimental value $10.7\times 10^{-4}$m$\Omega$cm/K for
$x=0.15$ \cite{Ni-Non-Fermi3}. This slope is expected to increase and
approach the experimental value when the CDW fluctuations neglected in this study are taken into account, since the $V^{\rm FLEX}$ is enhanced by $\chi^c$ near the CDW QCP.

To evaluate the three-dimensionality of the present model, we
introduce a magnification factor $r_z$ for the inter-layer hopping integrals.
$r_z=1$ correspond to the original 3D model. The three-dimensionality is
enlarged by $r_z>1$.
Figure \ref{fig4}(b) shows $T$ dependences of $\rho$ for
$r_z=1,2,2.5$ at $I_{\rm Nd}=20$eV. At low temperatures, $\rho\propto T^{0.84}$ for $r_z=1$, $\rho\propto T^{1.11}$ for $r_z=2$, and $\rho\propto
T^{1.42}$ for $r_z=2.5$ are obtained by the least-squares fittings for
$8\leq T\leq 15$meV.
In the case of $r_z=2.5$, $T$ dependence of $\rho$ is very close to
$\rho\propto T^{1.5}$, which corresponds to the behavior near the QCP of
3D systems according to the SCR theory
\cite{moriya-takahashi,ueda-moriya,moriya-ueda}.

\begin{figure}[!htb]
\includegraphics[width=.99\linewidth]{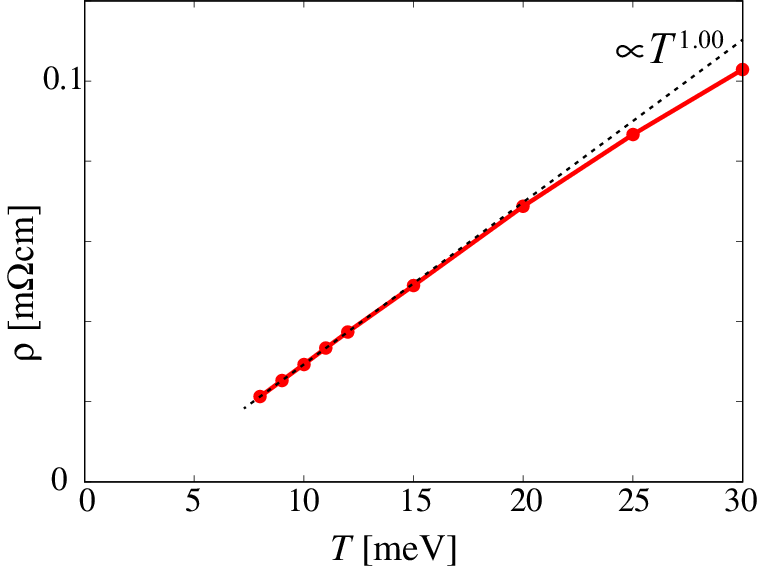}
\caption{
 $T$ dependence of $\rho$ for $I_{\rm Nd}=20$eV in the 2D model.
}
\label{fig-2D}
\end{figure}

In order to compare with the 2D system, we employ the 2D
model with only the $k_z=0$ plane for $r_z=1$. Figure \ref{fig-2D} shows
$T$-linear $\rho$ obtained in the 2D model for $U=2.6$eV, which corresponds to the behavior at the 2D QCP in the SCR theory.
As shown in Fig. \ref{fig4}(b), $T$ dependence of $\rho$ for $r_z=1$ is approximately $\rho\propto T$, which
reflects the behavior near the QCP of 2D systems.
This fact confirms that the electron state in the present 3D model for
$r_z=1$ is
a quasi-2D.
  These results are similar to CeCoIn$_5$, which has a 3D crystal
  structure but exhibits $\rho\propto T$ \cite{Ce-trans,Onari-Ce}.
  Thus, both electron states of $R$NiO$_2$ and CeCoIn$_5$ are quasi-2D.

\begin{figure}[!htb]
\includegraphics[width=.99\linewidth]{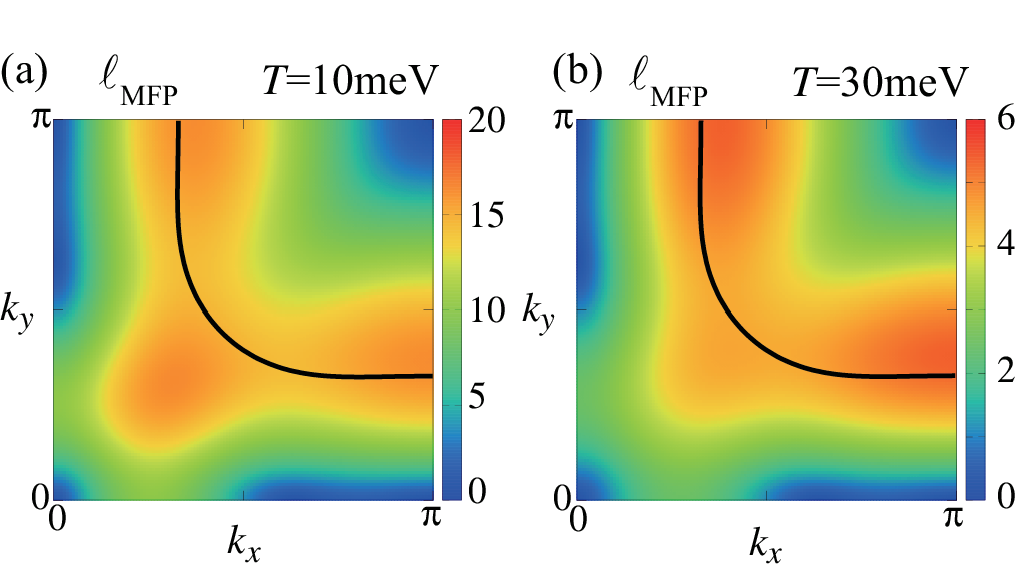}
 \caption{
(a) $\k$ dependence of mean-free path $\ell_{\rm MFP}$ on $k_z=0$ plane at $T=10$meV, and (b)
 that at $T=30$meV.
 Black line represents the FS.}
\label{fig-MFP}
\end{figure}

Here, we verify that the quasi-particle picture is valid in the present
study. Figure \ref{fig-MFP} shows the obtained mean-free path $\ell_{\rm
MFP}$ for Ni $d_{x^2-y^2}$ orbital at $T=10$meV, $30$meV in units of
lattice constant. $\ell_{\rm
MFP}$ is given as
\begin{equation}
 \ell_{\rm
MFP}(\k)=\frac{v^{b=1}_{\k}}{-{\rm Im}\Sigma^{b=1}_{\rm tot}(\k,0)}.
\end{equation}
On the FS, $\ell_{\rm MFP}\gg 1/k_{\rm
F}$, where $k_{\rm F}\sim 1$ is the Fermi wavenumber, is satisfied for $T\le
30$meV. In addition, the obtained slope of resistivity is consistent with
that in experiments. These facts indicate that the quasi-particle picture is valid for $T\le
30$meV in the present study. However, the quasi-particle picture may not be applicable at
higher temperatures ($T>30$meV).

\begin{figure}[!htb]
\includegraphics[width=.9\linewidth]{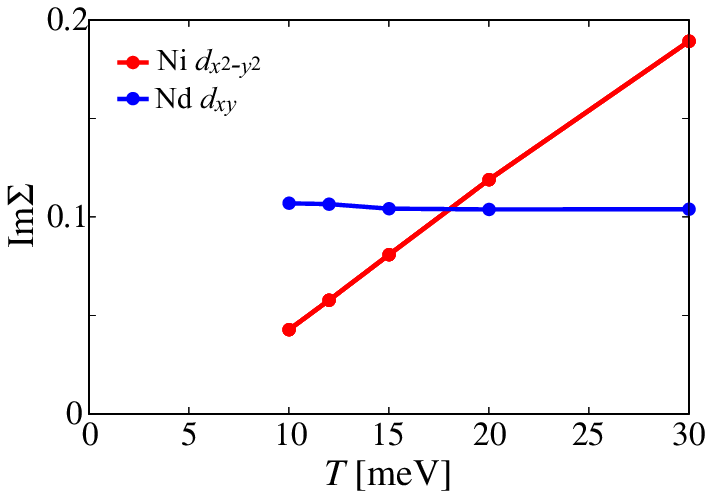}
 \caption{
 $T$ dependences of self-energy Im$\Sigma$ averaged on FS for orbitals 1
 and 3.
 }
\label{fig5}
\end{figure}

Here, we discuss the obtained $T$ dependence of $\rho$, which is proportional
to $T$ at low
temperatures, while it is suppressed and saturated at high temperatures.
Considering the orbital decomposition of the resistivity, we obtain
\begin{equation}
\rho=\left(\frac{1}{\rho_{\rm Ni}}+\frac{1}{\rho_{\rm Nd}}\right)^{-1}. \label{rho-tot}
\end{equation}
Here, we denote
\begin{equation}
\rho_{\rm Ni}\propto {\rm Im}\Sigma_{\rm Ni},\quad \rho_{\rm Nd}\propto {\rm
 Im}\Sigma_{\rm Nd},
\end{equation}
 where $\Sigma_{\rm Ni}$ and $\Sigma_{\rm Nd}$ are self-energies of orbital 1 and
that of orbital 2(3), respectively.
As shown in Fig. \ref{fig5}, $1/{\rm Im}\Sigma_{\rm
Ni}$ is dominant over $1/{\rm Im}\Sigma_{\rm Nd}$ at low temperatures,
since ${\rm Im}\Sigma_{Ni}\propto T$ is induced by the spin
fluctuation, and ${\rm Im}\Sigma_{Nd}$ by the impurity effect is
independent of $T$.
In contrast, $1/{\rm Im}\Sigma_{\rm
Nd}$ is dominant over $1/{\rm Im}\Sigma_{\rm Ni}$ at high temperatures.
Thus, $T$ dependence of $\rho$ is $\rho\sim \rho_{\rm Ni}\propto T$ at
low temperatures, and $\rho\sim \rho_{\rm Nd}$ at high temperatures.

\begin{figure}[!htb]
\includegraphics[width=.8\linewidth]{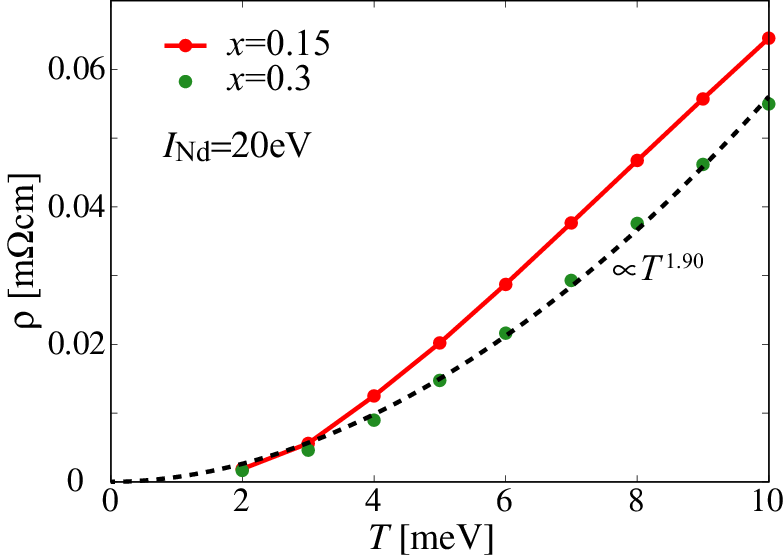}
\caption{
 Low $T$ dependence of $\rho$ for $I_{\rm Nd}=20$eV at $x=0.15,0.3$.
 Dashed line $\rho\propto T^{1.90}$ at $x=0.3$ is obtained by the least-squares fittings.
}
\label{low-T}
\end{figure}

In the following, we examine the doping $x$ dependence of $\rho$.
Low $T$ dependences of $\rho$ at $x=0.15,0.3$ are shown in Fig. \ref{low-T}.
At $x=0.3$, $\rho\propto T^{1.90}$ is obtained by the least-squares
fittings for $T<10$meV, which is consistent with the Fermi-liquid-like
behavior observed experimentally as
$\rho\sim T^2$ at $x\sim 0.3$ \cite{Ni-Non-Fermi3}.
This behavior is due to the reduction of spin fluctuations with
increasing hole doping $x$.
In contrast, at $x=0.15$, $T$-linear $\rho$ remains for $T\gtrsim
3$meV, which is also consistent with experimental results
\cite{Ni-Non-Fermi1,Ni-Non-Fermi2,Ni-Non-Fermi3}.
However, the obtained $T$ dependence of $\rho$ deviates from linear for $T\lesssim
2$meV, since the $x=0.15$ does not correspond to the QCP of the SDW.
The temperature range where the $T$-linear $\rho$ emerges is expected
to expand to lower temperatures when the CDW fluctuations are taken into
account. Analyzing the effect of CDW fluctuations on resistivity is an important future problem.

\begin{figure}[!htb]
\includegraphics[width=.9\linewidth]{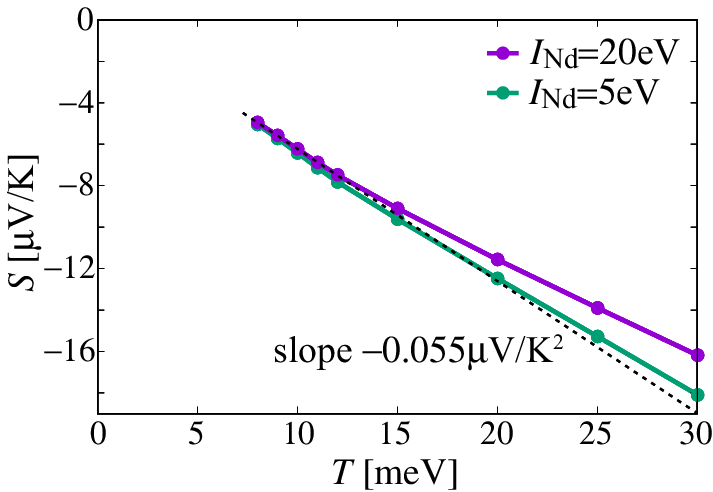}
 \caption{
$T$ dependence of $S$ for $I_{\rm Nd}=5,20$eV.
 }
\label{fig6}
\end{figure}

Finally, we analyze the Seebeck coefficient $S$.
Figure \ref{fig6} shows the $T$ dependence of $S$ for $I_{\rm
Nd}=5,20$eV.
At low temperatures, $T$-linear behavior of $S$ is obtained, which is
consistent with experimental results \cite{Ni-S}.
The slope $-0.055\mu$ V/K$^2$ for $I_{\rm
Nd}=20$eV is very close to the experimental value
$-0.050\mu$ V/K$^2$ in triple-layer and quintuple-layer nickelates \cite{Ni-S}.
This behavior of $S$ is similar to that obtained in the model for
YBCO cuprates \cite{S-kontani}.

 \section{Conclusion}
We studied the origin of non-Fermi-liquid $T$-linear resistivity in
infinite-layer nickelate Nd$_{0.85}$Sr$_{0.15}$NiO$_2$ based on a 3D
tight-binding model within the framework of the quasi-particle picture. The self-energies, calculated by the FLEX approximation for the
Ni orbital and the T-matrix approximation for an impurity
   effect on the Nd orbitals, are taken into account.
We found that the spin fluctuations induce the $T$-linear resistivity at low temperatures despite their 3D nature.
Thus, this system is regarded as quasi-2D, similar to CeCoIn$_5$.
We also found that the obtained Seebeck coefficient is negative and 
$T$-linear, which is consistent with experiments in triple-layer and quintuple-layer nickelates.

\begin{acknowledgment}
We are grateful to 
H. Kontani and
 Y. Yamakawa
for valuable discussions.
This work was supported by JSPS KAKENHI Grant Number JP23H03299.
\end{acknowledgment}

\appendix
\section{Details of the Present Model}
Here, we explain the details of the present model. 
For Ni $d_{x^2-y^2}$ orbital, we employ the intralayer next-nearest-neighbor
hopping $t_2$, the intralayer third-nearest one $t_3$, 
interlayer nearest one $t_z$, and next-nearest-interlayer one $t_{z2}$ to
$t_2/t_1=-1/3$, $t_3/t_1=0.2$, $t_z/t_1=2/3$, and  $t_{z2}/t_1=-0.165$ ($t_1=-0.376$eV is the nearest-neighbor
hopping), respectively. In addition, we introduce the energy shift
$-0.25$eV of Ni $d_{x^2-y^2}$ orbital to reproduce the small Fermi
pocket of Nd orbitals given by DFT+DMFT calculation \cite{Ni-first-principle6}. The hole-doping $x$ is introduced by the rigid-band
shift for all three bands.

\end{document}